\documentclass[preprint,showpacs,showkeys,preprintnumbers,amsmath,amssymb,prb,superscriptaddress]{revtex4}
\usepackage{bm}
\usepackage{amsmath}
\usepackage{graphicx}
\usepackage{amssymb}
\usepackage{ulem}
\usepackage{epsfig} 
\usepackage{dcolumn}
\usepackage[usenames]{color}

\begin{document}

\title{Set-based corral control in stochastic dynamical systems:\\
 Making almost invariant sets more invariant}

\author{Eric Forgoston\footnote{Corresponding author: eric.forgoston@montclair.edu}}

\affiliation{Department of Mathematical Sciences, Montclair State University, 1 Normal Avenue, Montclair, NJ 07043, USA}

\author{Lora Billings}

\affiliation{Department of Mathematical Sciences, Montclair State University, 1 Normal Avenue, Montclair, NJ 07043, USA}

\author{Philip Yecko}

\affiliation{Department of Mathematical Sciences, Montclair State University, 1 Normal Avenue, Montclair, NJ 07043, USA}

\author{Ira B.~Schwartz}

\affiliation{Nonlinear Systems Dynamics Section, Plasma Physics Division, Code 6792, U.S. Naval Research Laboratory, Washington, DC 20375, USA}

\begin{abstract}
We consider the problem of stochastic prediction and control in a time-dependent stochastic environment, such as the ocean, where escape from an almost invariant region occurs due to random fluctuations. We determine high-probability control-actuation sets by computing regions of uncertainty, almost invariant sets, and Lagrangian Coherent Structures. The combination of geometric and probabilistic methods allows us to design regions of control that provide an increase in loitering time while minimizing the amount of control actuation. We show how the loitering time in almost invariant sets scales exponentially with respect to the control actuation, causing an exponential increase in loitering times with only small changes in actuation force. The result is that the control actuation makes almost invariant sets more invariant.
\end{abstract}



\maketitle

{\bf Prediction and control of the motion of an object in time-dependent
and stochastic environments is an important and fundamental problem
in nonlinear dynamical systems. One of the main goals of control is
the design of a theory that can take unstable states and render them
stable. For example, small perturbations at the base of an inverted
pendulum will stabilize the inverted state. Noise poses a greater
problem for deterministically controlled states, in that stochastic
effects destabilize the states as well as their neighborhoods. Therefore,
control theory of stochastic dynamical systems may be addressed by
examining the change in stability of certain sets. 

We present a variety of geometric and probabilistic
set-based methods that enable one to compute controllable sets. For
a particle moving under the influence of deterministic and stochastic
forces with no control, these sets determine regions which are unstable
in the sense that the particle will leave the set after a sufficiently long
time. Controls added to particle dynamics to increase the time to escape (or
loitering time) have a strong dependence on the probabilistic and geometric set
characteristics. The determination of controls and associated sets
allow for an increase in the amount of time the particle can loiter
in a particular region while minimizing the amount of control actuation. 

Our theoretical analysis shows how an increase in the strength of
the control force leads to a decrease in the probability that an object
will escape from the control region. In fact, we have found that small
changes in the control actuation force have an exponential effect
on the loitering time of the object. Additionally, we show how the
exponential increase in escape times from the controlled sets is related
to the problem of noise-induced escape from a potential well.
}

\section{Introduction}

Ocean circulation impacts weather, climate, marine fish and mammal populations, and contaminant transport, making ocean dynamics of great industrial, military and scientific interest. A major problem in the field of ocean dynamics involves forecasting, or predicting, a variety of physical quantities, including temperature, salinity and density. By fusing recently measured data with detailed flow models, it is possible to achieve improved prediction. Therefore, forecasts occur with increased accuracy~\cite{HarlimOYKH05}. As a result, ocean surveillance may be improved by incorporating the continuous monitoring of a region of interest.

Researchers have used surface drifters and submerged floats to acquire data for many years. More recently, sensing platforms such as autonomous underwater gliders~\cite{webbsijo01,shermandov01,eriksenolwlsbc01} have been developed. The gliders can operate in both littoral (coastal) and deep-ocean regimes, and may be used for data acquisition, surveillance and reconnaissance. One drawback in the use of gliders involves their limited amount of total control actuation due to energy constraints, such as short battery life. For applications such as regional surveillance, energy constraints may be alleviated by taking advantage of the dynamical flow field structures and their respective body forces. 

Autonomous underwater gliders are subjected to drift due to hydrodynamic forces. This drift can be extremely complicated since the velocity fields found in the ocean are aperiodic in time and feature a complex spatial structure. Instead of constantly reacting to the drift (and thereby expending energy), one can minimize the glider's energy expenditure by taking advantage of the underlying structure found in geophysical flows. However, in order to harness the ocean forces to minimize energy expenditure during control actuation, one needs to analyze the structures from the correct dynamical viewpoint. 
 
The potential for dynamical systems tools to shed light on complex, even turbulent, flow fields such as ocean dynamics, has been understood for decades. Work in this area has intensified in the past decade with new focus placed on the improved Lagrangian perspective that dynamical systems approaches may provide, especially for complex, aperiodic flows in which the traditional Eulerian perspective on the flow is unhelpful or even misleading. During boundary layer separation, for example, sheet-like flow structures coincide with fluid particles being ejected from the wall region, a technologically important feature that an Eulerian framework fails to capture in unsteady cases, but which has a clear Lagrangian signature that can be identified using finite-time Lyapunov exponents~\cite{WeldonEtAl_JFM2008}. In liquid jet breakup, a similar ejection process occurs during primary atomization, but a limited Lagrangian perspective is provided by the liquid-gas interface, revealing liquid sheets and ligaments that precede droplet formation~\cite{VillMarm}. The development of these critical fluid sheets and ligaments can be traced to unsteady, finite-amplitude global flow structures~\cite{BoeckEtAl_TCFD}. Dynamical systems tools thus provide a new approach to the characterization and control of these important flows. Geophysical flows offer another example where an Eulerian framework is ineffective in the diagnosis of large Lagrangian structures and the measurement of transport. For prediction and control of particle dynamics in large surveillance regions of interest, Lagrangian structures of geophysical flows need to be characterized in both deterministic and stochastic settings.

\begin{figure}
\begin{centering}
\includegraphics[width=8.5cm,height=8.5cm]{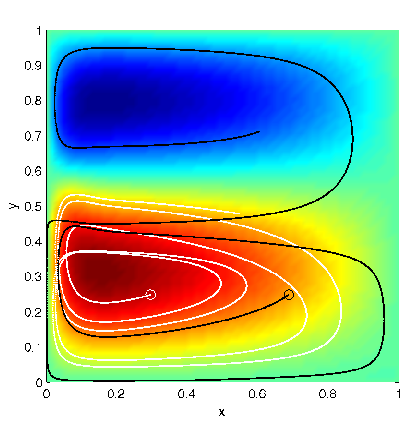} 
\par\end{centering}

\caption{\label{fig:ocean} (Color online) Streamfunction (color) and two particle trajectories within a single-layer quasi-geostrophic ocean model (see Appendix~\ref{sec:ocean_model}) subject to low-amplitude periodic forcing whose mean gives a steady double-gyre/western boundary current flow solution. Initial points on the trajectory are identified with open circles. 
}
 \end{figure}

The field of Geophysical Fluid Dynamics (GFD) involves the study of large-scale fluid flows that occur naturally. GFD flows are, by nature, aperiodic and stochastic. The data sets describing them are usually finite-time and of low resolution. Established tools of dynamical systems have proven to be less effective in these cases: while providing some insight, they cannot provide realistic or detailed flow field data relevant to the trajectories of tracer particles. For this, dynamical systems tools can be applied to fluid flows in an alternative way, by interpreting the Eulerian flow field ${\bf u}({\bf x},t)$ as a dynamical system ${\bf x}'={\bf u}({\bf x},t)$ describing the trajectories of tracer particles. The phase space and real space are identical here and due to the incompressibility condition $\nabla\cdot{\bf u}=0$, the resulting dynamical system is conservative. This {}``chaotic advection'' approach originated with Aref~\cite{Aref} and demonstrated that even simple laminar two-dimensional (2D) periodic and three-dimensional (3D) steady flows~\cite{ABC} could lead to complex, chaotic particle trajectories. Figure~\ref{fig:ocean} illustrates an example of how particles in a periodic flow may exhibit unexpected trajectories. In this figure, a single-layer quasi-geostrophic beta plane is being driven by a bimodal wind-stress with a small amplitude periodic perturbation. Details of this ocean model can be found in Appendix~\ref{sec:ocean_model}. 

In the last two decades, this approach has led to new tools including the
study of transport by coherent structures~\cite{Anto}, lobe
dynamics~\cite{Wig}, distinguished trajectories~\cite{Ide}, and global bifurcations~\cite{BerMea97,NaLu01}. Even though the transport controlling structures in GFD flows are inherently complicated and unsteady, their understanding is necessary to the design of glider controls. To overcome this obstacle we combine a set of dynamical systems tools that have proven effective in higher dimensions~\cite{Meucci} and stochastic problems~\cite{forsch09,fobisc09}.
 
In this paper, we will consider a well-known driven double-gyre flow as an example to illustrate our prediction/control framework~\cite{YanLiu94}. This model can be thought of as a simplified version of the double-gyre shown in Fig.~\ref{fig:ocean} which is a solution to a realistic quasi-geostrophic ocean model. It should be noted that our methods are general and may be applied to any flow of interest. The goal of our approach is the production of a complete picture of particle trajectories and tracer lingering times that enables one to design a control strategy that limits tracers from switching between gyres. 

The techniques we will use to analyze the dynamical systems are based on both
deterministic and stochastic analysis methods, and reveal different structures
depending on the system under examination. In the deterministic case, the
Lagrangian Coherent Structures reveal much about transport and information
related to the basin boundaries. They also pinpoint
regions of local sensitivity in the phase space of interest. Since basin
boundaries are most sensitive to initial conditions, uncertainties in the data
near the boundaries generate obstructions to predictability. Therefore, highly
uncertain regions in phase space may be revealed by computing local
probability densities of uncertain regions in the deterministic case, but
using noisy initial data. Finally, in the time-dependent
stochastic case, we can describe which sets contain very long term, albeit
finite, trajectories. The sets are almost invariant due to the stochastic
forcing on the system, which causes random switching between the almost
invariant sets. The tools we use here are based on the stochastic
Frobenius-Perron operator theory. Once the full structure of almost invariant
sets is identified along with regions of high uncertainty, control strategies
may be designed to maintain long time trajectories within a given region with
minimal actuation.  In Fig.~\ref{fig:ocean}, one can see that one
  particle's trajectory remains in the lower gyre for a long period of time, while the other particle escapes from the lower gyre to the upper gyre.  The
tools we will develop and outline in this article will enable one to know if
and when
a control force must be actuated to prevent the particle from escaping.
 
The layout of the paper is as follows. In Sec.~\ref{sec:themodel} we present
the stochastic double-gyre system and examine the deterministic dynamical
features. In Sec.~\ref{sec:ftle}, we show how to use the finite-time Lyapunov
exponents to describe transport, and we show in Sec.~\ref{sec:unc} how to
quantify uncertainty regions. We then turn to the stochastic system, and
describe in Sec.~\ref{sec:ais} how to compute almost invariant
sets. Section~\ref{sec:control} contains a discussion of our corral control strategy, and Sec.~\ref{sec:conc} contains the conclusions and discussion. 

\section{The model}

\label{sec:themodel} A simple model of the wind-driven double-gyre flow
is provided by \begin{subequations} \begin{flalign}
 & \dot{x}=-\pi A\sin(\pi f(x,t))\cos(\pi y)-\alpha x +\eta_{1}(t),\label{e:toygyre_a}\\
 & \dot{y}=\pi A\cos(\pi f(x,t))\sin(\pi y)\frac{df}{dx}-\alpha y +\eta_{2}(t),\label{e:toygyre_b}\\
 & f(x,t)=\epsilon\sin(\omega t +\psi)x^{2}+(1-2\epsilon\sin(\omega t +\psi))x.\label{e:toygyre_c}\end{flalign}
 \end{subequations} 
When $\epsilon =0$, the double-gyre flow is time-independent, while for $\epsilon\neq 0$, the gyres undergo a periodic expansion and contraction in the $x$-direction. In Eqs.~(\ref{e:toygyre_a})-(\ref{e:toygyre_c}), $A$ approximately determines the amplitude of the velocity vectors, $\omega /2\pi$ gives the oscillation frequency, $\epsilon$ determines the amplitude of the left-right motion of the separatrix between the two gyres, $\psi$ is the phase, $\alpha$ determines the dissipation, and $\eta_{i}(t)$ describes a stochastic white noise with mean zero and standard deviation $\sigma=\sqrt{2D}$, for noise intensity $D$. This noise is characterized by the following first and second order statistics: $\left <\eta_{i}(t)\right >=0$, and $\left <\eta_{i}(t)\eta_{j}(t')\right >=2D\delta_{ij}\delta(t-t')$ for $i=1,2$. In the rest of the article we shall use the following parameter values: $\alpha=0.005$, $A=0.1$, $\epsilon=0.15$, and $\omega=2\pi /20$. We consider the dynamics restricted to the domain $\{(x,y)\,|\,0\le x\le2$ and $0\le y\le1\}$. An unforced, undamped autonomous version of the double-gyre model was studied by Rom-Kedar, et.al.~\cite{Rom-Kedar}, and an undamped system with different forcing was studied by Froyland and Padberg~\cite{Froyland09}.

\begin{figure}
\begin{centering}
\includegraphics[width=8.5cm,height=4.25cm]{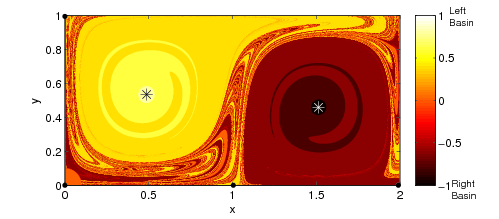} 
\par\end{centering}

\caption{\label{fig:basin} (Color online) The basin Poincar\'{e} map for the deterministic part of Eqs.~(\ref{e:toygyre_a})-(\ref{e:toygyre_c}) at phase $\psi=0$. The locations of the attracting fixed points corresponding to periodic orbits are denoted by the stars. The coloring represents the convergence rate to the attractors. As shown by the color bar, positive values converge to the left basin and negative values converge to the right basin. The largest magnitude values converge the fastest. Four saddle fixed points are located within the boundaries of the domain, and are denoted by large dots. The remaining two are located at $(0.995,1.005)$ and $(2.010,0.995)$.}

\end{figure}

Prior to defining the control of certain sets of the stochastic system, we
first describe the important dynamical features of the deterministic part of
Eqs.~(\ref{e:toygyre_a})-(\ref{e:toygyre_c}). There are two attracting
periodic orbits, which correspond to two fixed points of the Poincar\'{e} map
defined by sampling the system at the forcing period. For each fixed point,
there corresponds a left and a right basin of attraction. Local analysis on
the Poincar\'{e} section about the fixed points reveals the attractors to be
spiral sinks, which generate the global double-gyre.  A representative basin map at the phase $\psi=0$ is shown in Fig.~\ref{fig:basin}. One can see the complicated basin boundary structure in which the basins of attraction are intermingled, a signature of the existence of a fractal basin boundary. There are also several unstable (saddle) periodic orbits corresponding to fixed points that lie along or close to the domain boundary.

Due to the intermingling of the basin boundaries, one can expect that small
perturbations, or uncertainties, in initial conditions near the basin boundary
will generate large changes in dynamical behavior. This can occur
deterministically, or when noise is added to the system. Therefore, we
quantify regions of uncertainty in phase space in both deterministic and
stochastic settings in Sections IV and V respectively. 

\section{Finite-time Lyapunov exponents}

\label{sec:ftle} One method that can be used to understand transport and which
quantifies localized sensitive dependence to initial conditions in a given
fluid flow involves the computation of finite-time Lyapunov exponents
(FTLE). In a deterministic setting, the FTLE also gives an explicit measure of
phase space uncertainty. Given a dynamical system, one is often interested in
determining how particles that are initially infinitesimally close behave as time $t\to\pm\infty$. It is well-known that a quantitative measure of this asymptotic behavior is provided by the classical Lyapunov exponent~\cite{guchol86}. In a similar manner, a quantitative measure of how much nearby particles separate after a specific amount of time has elapsed is provided by the FTLE.

In the early 1990s, Pierrehumbert~\cite{pie91} and Pierrehumbert and Yang~\cite{pieyan93} characterized atmospheric structures using FTLE fields. In particular, their work enabled the identification of both mixing regions and transport barriers. Later, in a series of papers published in the early 2000s, Haller~\cite{hall00,hall01,hall02} introduced the idea of Lagrangian Coherent Structures (LCS) in order to provide a more rigorous, quantitative framework for the identification of fluid structures. Haller~\cite{hall02} proposed that the LCS be defined as a ridge of the FTLE field, and this idea was formalized several years later by Shadden, Lekien and Marsden~\cite{shlema05}. When computing the FTLE field of a dynamical system, these LCS, or ridges, are seen to be the structures which have a locally maximal FTLE value.

Although the FTLE/LCS theory can be extended to arbitrary dimension~\cite{leshma07}, in this article we consider a 2D velocity field ${\bm{v}}:\mathbb{R}^{2}\times I\rightarrow\mathbb{R}^{2}$ given by the deterministic part of Eqs.~(\ref{e:toygyre_a})-(\ref{e:toygyre_c}) which is defined over the time interval $I=[t_{i},t_{f}]\subset\mathbb{R}$ and the following system of equations: \begin{subequations} \begin{flalign}
 & \dot{{\bm{z}}}(t;t_{i},{\bm{z}}_{0})={\bm{v}}({\bm{z}}(t;t_{i},{\bm{z}}_{0}),t),\label{e:xdot}\\
 & {\bm{z}}(t_{i};t_{i},{\bm{z}}_{0})={\bm{z}}_{0},\label{e:xIC}\end{flalign}
 \end{subequations} where ${\bm{z}}=(x,y)^{T}\in\mathbb{R}^{2}$,
${\bm{z}}_{0}\in\mathbb{R}^{2}$, and $t\in I$.

As previously stated, the trajectories of this dynamical system in the infinite time limits can be quantified with the system's Lyapunov exponents. If one restricts the Lyapunov exponent calculation to a finite time interval, the resulting exponents are the FTLE. In practice, the FTLE computation involves consideration of nearby initial conditions and the determination of how the trajectories associated with these initial conditions evolve in time. Therefore, the FTLE provides a local measure of sensitivity to initial conditions and measures the growth rates of the linearized dynamics about the trajectories. Since the details of the derivation of the FTLE~\cite{hall00,hall01,hall02,shlema05,leshma07,brawig09} as well as applications that employ the FTLE~\cite{tachha10,eldcho10,luyafa10} have appeared in the literature, we shall only briefly summarize the procedure.

The solution of the dynamical system given by Eqs.~(\ref{e:xdot})-(\ref{e:xIC}) from the initial time $t_{i}$ to the final time $t_{i}+T$ can be viewed as the flow map $\phi_{t_{i}}^{t_{i}+T}$ which is defined as follows: 
\begin{equation} \phi_{t_{i}}^{t_{i}+T}:{\bm{z}}_{0}\mapsto\phi_{t_{i}}^{t_{i}+T}({\bm{z}}_{0})={\bm{z}}(t_{i}+T;t_{i},{\bm{z}}_{0}).\label{e:map}
\end{equation} We consider an initial point located at ${\bm{z}}$ at $t_{i}=0$ along with a perturbed point located at ${\bm{z}}+\delta{\bm{z}}(0)$ at $t_{i}=0$. Using a Taylor series expansion, one finds that \begin{equation} \delta{\bm{z}}(T)=\frac{d\phi_{t_{i}}^{t_{i}+T}({\bm{z}})}{d{\bm{z}}}\delta{\bm{z}}(0)+\mathcal{O}(||\delta{\bm{z}}(0)||^{2}).\end{equation} Dropping the higher order terms, the magnitude of the linearized perturbations is given as 
\begin{equation} ||\delta{\bm{z}}(T)||=\sqrt{\left<\delta{\bm{z}}(0),\Delta\right>},
\end{equation} where $\Delta$ is the right Cauchy-Green deformation tensor and is given as follows: 
\begin{equation}
\Delta({\bm{z}},t_{i},T)=\left(\frac{d\phi_{t_{i}}^{t_{i}+T}({\bm{z}}(t))}{d{\bm{z}}(t)}\right)^{*}\left(\frac{d\phi_{t_{i}}^{t_{i}+T}({\bm{z}}(t))}{d{\bm{z}}(t)}\right),
\label{e:delta}
\end{equation}
 with {*} denoting the adjoint. Then the FTLE can be defined as 
\begin{equation}
\sigma({\bm{z}},t_{i},T)=\frac{1}{|T|}\ln{\sqrt{\lambda_{{\rm max}}(\Delta)}},
\label{e:sigma}
\end{equation}
where $\lambda_{{\rm max}}(\Delta)$ is the maximum eigenvalue of $\Delta$.

For a given ${\bm{z}}\in\mathbb{R}^{2}$ at an initial time $t_{i}$, Eq.~(\ref{e:sigma}) gives the maximum finite-time Lyapunov exponent for some finite integration time $T$ (forward or backward), and provides a measure of the sensitivity of a trajectory to small perturbations.

The FTLE field given by $\sigma({\bm{z}},t_{i},T)$ can be shown to exhibit ridges of local maxima in phase space. The ridges of the field indicate the location of attracting (backward time FTLE field) and repelling (forward time FTLE field) structures. In 2D space, the ridge is a curve which locally maximizes the FTLE field so that transverse to the ridge one finds the FTLE to be a local maximum.

\begin{figure}
\begin{centering}
\includegraphics[width=8.5cm,height=4.25cm]{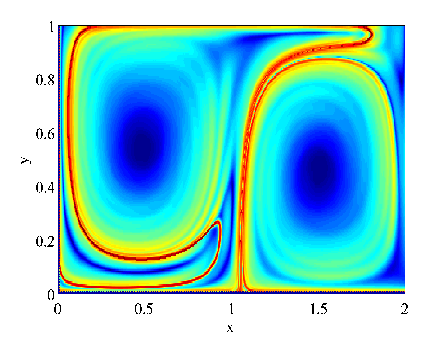} 
\par\end{centering}

\caption{\label{fig:FTLE} (Color online) Forward FTLE flow field computed using the deterministic form of Eqs.~(\ref{e:toygyre_a})-(\ref{e:toygyre_c}) and shown at phase $\psi=0$. The integration time is $T=20$ with an integration step size of $t=0.1$ and a grid resolution of $0.005$ in both $x$ and $y$. }

\end{figure}

Figure~\ref{fig:FTLE} shows a snapshot taken at phase $\psi=0$ of the forward time FTLE field calculated using the deterministic form of Eqs.~(\ref{e:toygyre_a})-(\ref{e:toygyre_c}) for a finite integration time $T=20$. One can see in Fig.~\ref{fig:FTLE} that there are ridges (in red) of locally maximal FTLE values. These ridges or LCS effectively separate the phase space into distinct dynamical regions. For the deterministic system, a particle placed in one of these distinct regions will remain in that region as the system evolves in time. Notice also that the red ridges appear to be a subset of the basin boundary set separating the two basins of attraction shown in Fig. \ref{fig:basin}.

In the stochastic system, the noise acts continuously on a particle placed in
one of these distinct basins, and therefore, it is possible that the particle
will cross the LCS and move into the other basin. Even though we find the LCS
using the deterministic system, the location of these structures are a
valuable tool in understanding how a particle escapes from the basin in which
it is initially placed. Coupled with the ideas of uncertain sets and almost
invariant sets described below, we will use the LCS information provided by
the FTLE field to determine appropriate regions of control. In this way, it is
possible to increase the loitering times of a particle in a basin while minimizing the amount of control actuation. 

\section{Uncertain Sets}

\label{sec:unc}

In this section, we describe a method that measures the fraction of uncertainty for regions in phase space. Nonlinear systems that possess multiple attractors typically can be extremely sensitive with respect to the choice of initial conditions. That is, very small changes, or uncertainty, in the initial conditions can lead to different attractors~\cite{GrebigiMcDonald1983,McDonaldGrebogi1985,LaiGrebogi1994}. Sensitivity here is measured in the asymptotic time limit. When applied to high-dimensional attractors, long-time sensitivity is measured with respect to parameter sensitivity~\cite{SchwartzWood1999}. In contrast to the asymptotic definition, we measure uncertainty in phase space with respect to perturbations in initial data by computing exponential changes in distances over short time intervals.

\begin{figure}
\begin{centering}
\includegraphics[width=8.5cm,height=4.25cm]{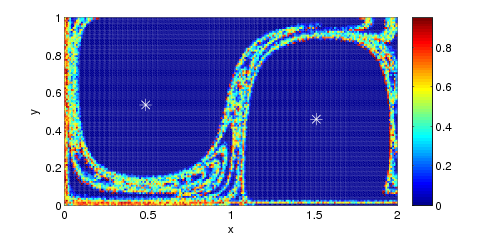} 
\par\end{centering}

\caption{\label{fig:uncertain} (Color online) The uncertainty measure for the
  deterministic part of Eqs.~(\ref{e:toygyre_a})-(\ref{e:toygyre_c}) for the
  phase $\psi=0$. A 200 x 100 regular grid of initial
  conditions is used. The color represents the fraction of trajectories that diverge
  from the trajectory of the center of each ball tested. The computation was
  performed using balls of radius $\epsilon=0.02$ with twenty points chosen at random within each ball. The trajectories were computed until time $\tau=400$ using a threshold parameter of $0.8$.}

\end{figure}

The sensitivity of the deterministic system to the initial conditions is
quantified by defining the pointwise uncertainty measure. For a point in phase
space $(x_{0},y_{0})$ chosen at random, a ball of radius $\epsilon$ is
constructed about that point, $B_{\epsilon}(x_{0},y_{0})$. For $N$ randomly
chosen points in the ball $B_{\epsilon}(x_{0},y_{0})$, we record the number of
trajectories, $N_{\epsilon}(\tau,x_{0},y_{0})$, at time $\tau$, that diverge
beyond a certain distance from the trajectory of $(x_{0},y_{0})$. By diverge,
we mean that we test if the distance between the trajectories at a given time
is greater than a threshold parameter. The test indicates the fraction of
initial conditions in the ball that diverge beyond a certain distance in a
given finite time. If there are multiple attractors present, then as
$\tau\rightarrow\infty$, the uncertain points correspond to those points that
go to another basin. For finite $\tau$, as the radius of the ball is decreased, one obtains more accurate approximations of the basin boundary of the system. This is due to the fact that the dynamics are most sensitive with respect to the choice of initial conditions when the points straddle the basin boundary separating the basins of attraction.

Figure~\ref{fig:uncertain} shows a sample calculation of the uncertainty
measure for the fixed phase $\psi=0$. In this calculation, a 200 x 100 regular
grid is used where each grid point coincides with the center of a ball of
radius 0.02. By comparing the FTLE field shown in Fig.~\ref{fig:FTLE} with the
uncertainty measure shown in Fig.~\ref{fig:uncertain}, one can see that the location of the FTLE ridge is contained within the uncertainty region. In addition, it should be noted that the uncertainty measure captures the complicated fractal structure of the basin boundary shown in Fig.~\ref{fig:basin}, as radius $\epsilon$ approaches zero. For the stochastic double-gyre, it is expected that the regions defined by the uncertain sets over all phases, $\psi\in[0,20)$ will be the most active regions in which to actuate if one wishes to increase the residence time of a particle in one particular basin. We now refine this notion in the discrete case by defining the actual sets which will be almost invariant in the stochastic case. One may think of the almost invariant sets as the complement of the most uncertain regions in phase space.

\section{Almost Invariant Sets}

\label{sec:ais}

As mentioned above, approximating the almost invariant sets is another method
that can quantify where particles loiter in phase space for a stochastic
system.  Computation of the almost invariant sets will be achieved
using the stochastic Frobenius-Perron (SFP) operator.  While the transition probabilities can be approximated for
   continuous time~\cite{Risken96}, we take advantage of the fact that the double-gyre system is
   a periodically forced system. 
 Therefore, the dynamics are sampled at a particular phase of the period.
For small noise, contributions to the probability along trajectories come
primarily from the difference between the tangent vector and the vector
field. Averaging over a period of the drive allows one to sample the noise
periodically.  In doing this,
we assume that the noise is not in the tail of the probability distribution so
that there are no large, rare events.
 We therefore construct a Poincar\'{e} map with a specified noise
 distribution.  
This method of discrete sampling  quantitatively identifies the complement of
the uncertainty region as two almost invariant sets, which are associated with
the stable attractors of the associated deterministic system. Although we
present the machinery for a fixed phase, it is possible to extend the analysis
for all phases of $\psi$, allowing one to approximate the density over the full
forcing period. 

To be explicit, consider a stochastically perturbed map $F_{\eta}:\mathbb{R}^{n}\rightarrow \mathbb{R}^{n},\, x(t)\mapsto F(x(t))+\eta$, where $\eta$ is a random variable having the distribution $\nu(x)$. Let $F(x(t))$ represent the map on the Poincar\'{e} section which samples the flow at time intervals of length $\tau$, equivalent to the period of the steady states. The SFP operator for a normal distribution with mean zero and  standard deviation $\sigma$ is defined as~\cite{BillingsBS02}
\begin{equation}
P_{F}[\rho(x)]=\frac{1}{\sqrt{2\pi\sigma^{2}}}\int_{M}e^{-\frac{\Vert(x-F(y))\Vert^{2}}{2\sigma^{2}}}\rho(y)dy.\label{FPnormal}\end{equation}
 One approximates the density $\rho(x)$ by the finite sum of basis functions,
 $\rho(x)\simeq\sum_{i=1}^{N}c_{i}\phi_{i}(x)$, where
 $\phi_{i}(x)=\chi_{B_{i}}(x)$, and $\chi_{B}$ is an indicator defined on
 boxes $\{B_{i}\}_{i=1}^{N}$ covering the region $M$. The SFP operator is approximated by the $N\times N$ matrix, \begin{equation}
A_{ij}\equiv A(B_{i},B_{j})=(P_{F}[\phi_{j}],\phi_{i})=\int_{M}P_{F}[\phi_{j}(x)]\phi_{i}(x)dx\label{eq:Aij matrix}\end{equation}
 for $1\leq i,j\leq N$. Therefore, a transition matrix entry $A_{ij}$ value represents how mass, or measure, flows from cell $B_{i}$ to cell $B_{j}$. Details of the method can be found in Ref.~\onlinecite{bbs02}.

\begin{figure}
\begin{centering}
\includegraphics[height=5cm]{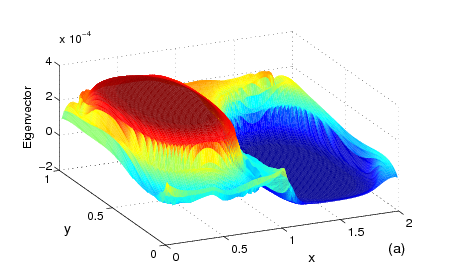}
\includegraphics[height=5cm]{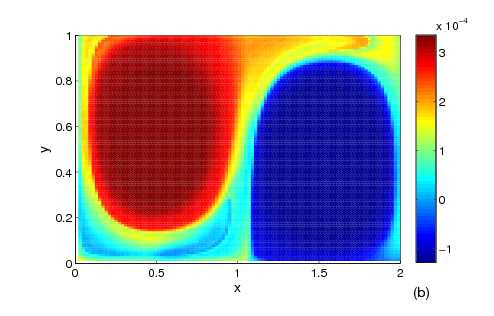} 
\end{centering}

\caption{\label{fig:almostinvariant} (Color online) The results of
  approximating the almost invariant sets for the stochastic system given by
  Eqs.~(\ref{e:toygyre_a})-(\ref{e:toygyre_c}) at the phase $\psi=0$. A 100 x
  100 grid of points is used with a standard deviation of $\sigma=0.05$. Figure \ref{fig:almostinvariant}(a) is the mapping of the second eigenvector of the reversible matrix back to the phase space. The almost invariant sets are represented by the red and the blue regions at the ends of the eigenvector range. Figure \ref{fig:almostinvariant}(b) is the contour plot of the surface, which shows the transition region that forms a barrier between the almost invariant sets.}

\end{figure}

From the transition matrix, a reversible Markov chain is constructed,
satisfying $\pi_{i}A_{ij}=\pi_{j}A_{ji}$ for all $i,j$. Such a condition
implies $\pi$ is an invariant (stationary) distribution of
$A$~\cite{Doob}. Since in general $A$ is not reversible as it is defined in
Eq.~(\ref{eq:Aij matrix}), a reversible Markov chain is constructed from the transition matrix. Let $R=(A+\hat{A})/2$, where $\hat{A}_{ij}=\frac{\pi_{j}}{\pi_{i}}A_{ji}$ and $\pi$ is an invariant probability density of $A$. The matrix $R$ now possesses detailed balance.

Since $R$ is reversible, it has the properties that its eigenvalues are real
and the eigenvectors are orthogonal. One can then compute a collection of sets
which are almost invariant by examining the eigenvectors of $R$. A gauge of
how much measure is ejected each iterate from the defined almost invariant
sets is given by the eigenvalues. The first few eigenvalues of $R$ will be
clustered near unity and their associated eigenstates will be a mixture of
almost invariant sets formed from the above basis elements. Since the first
eigenvector of $R$ is a vector of ones, the second eigenvector is used to identify two almost invariant sets emanating from basins of attraction. The fact that the eigenvalues cluster near unity means that very little mass is ejected from the basins each iterate, which is why the basins are defined as almost invariant states.

The results of the almost invariant sets computation are shown in
Fig.~\ref{fig:almostinvariant} for the Poincar\'{e} map of the double-gyre at
$\psi=0$. The second eigenvector can be visualized by a roughly piecewise
constant function, with a transition region connecting the two pieces. The
eigenvalue associated with this eigenvector is 0.9996. The higher level
represents the left basin (red) and the lower level represents the right basin
(blue). For this computation, a 100 x 100 grid of points is used with a standard deviation of $\sigma=0.05$. As expected, the complement set of the left and right almost invariant basins agrees approximately with the location of the uncertain region from Fig.~\ref{fig:uncertain}.

\section{Corral control in the presence of fluctuations}

\label{sec:control}

\begin{figure}
\begin{centering}
\includegraphics[width=8.5cm,height=4.25cm]{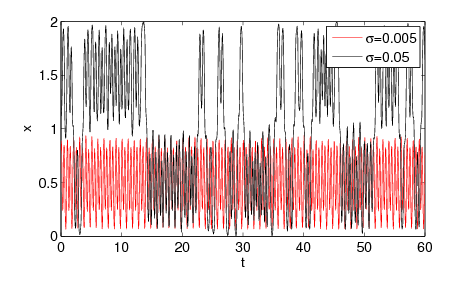} 
\end{centering}
\caption{\label{fig:switching} (Color online) A graph of the
  $x-$coordinate for two simulations of the stochastic system given by
  Eqs.~(\ref{e:toygyre_a})-(\ref{e:toygyre_c}). For continuous noise with
  standard deviation $\sigma=0.005$, one does not
    observe switching events (red).  However, for the
  larger standard deviation of $\sigma=0.05$, one can see
  frequent switching between the two basins (black).}

\end{figure}

Since we have now computed the location of the Lagrangian Coherent
  Structures and have approximated the most uncertain regions and almost
invariant regions of phase space, we can consider loitering, or residence
times, within a defined region. Moreover, we may use actuation to control trajectories in the system by increasing the loitering times within an almost
invariant set. To quantify the residence time of a particle in a basin, we
consider the double-gyre system with continuously added Gaussian noise. As the
standard deviation, $\sigma$, of the noise is increased, trajectories wander
in larger neighborhoods about the stable steady states and frequently switch
from one basin to another, as shown in Fig.~\ref{fig:switching}. These
basins are approximated by the almost invariant sets at a fixed phase, shown in Fig.~\ref{fig:almostinvariant}. 

The control strategy employs a monitoring ball (that we refer to as a control
region) that covers the almost invariant set with the center defined as the
left non-boundary steady state when time $t=0$. When a trajectory passes
beyond a threshold distance from the center $(x_{s},y_{s})$ of the control region, a radial force is turned on. This force moves the trajectory back towards the center of the region. This control strategy can be modeled as
\begin{subequations} 
\begin{flalign}
 & \dot{x}=-\pi A\sin(\pi f(x,t))\cos(\pi y)-\alpha x +\eta_{1}(t)-c(x-x_{s})\Theta(r(x,y)-r_{0}),\label{e:xdot_c}\\
 & \dot{y}=\pi A\cos(\pi f(x,t))\sin(\pi y)\frac{df}{dx}-\alpha y +\eta_{2}(t)-c(y-y_{s})\Theta(r(x,y)-r_{0}),\label{e:ydot_c}\\
 & f(x,t)=\epsilon\sin(\omega t +\psi)x^{2}+(1-2\epsilon\sin(\omega t +\psi))x,\\
 & r(x,y)=\sqrt{(x-x_{s})^{2}+(y-y_{s})^{2}}, \label{e:r_c}
\end{flalign}
\end{subequations} 
where $c$ is the magnitude of the control force, $r_0$ is the radial
threshold, and $\Theta$ is a Heaviside function. By adjusting $r_{0}$ and $c$,
it is possible to control the number of actuations and the effectiveness of
the control scheme.  We have designed this specific control strategy to take
advantage of the underlying flow structure in order to corral the particles
into the almost invariant set.  We therefore refer to our scheme as corral control.

\begin{figure}
\begin{centering}
\includegraphics[height=8cm]{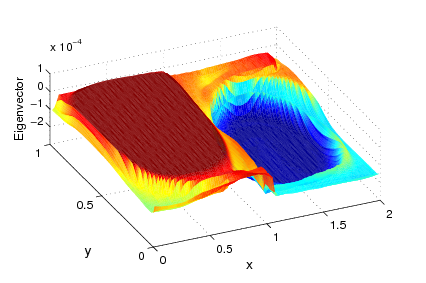} 
\end{centering}
\caption{\label{fig:almostinvariant_control} (Color online) The results of
  approximating the almost invariant sets for the stochastic system with
  control applied to the left basin given by
  Eqs.~(\ref{e:xdot_c})-(\ref{e:r_c}) at the phase $\psi=0$. A 100 x
  100 grid of points with a standard deviation of $\sigma=0.05$ is used. Similar to
  Figure \ref{fig:almostinvariant}(a), this graph is the mapping of the second
  eigenvector of the reversible matrix back to the phase space. The almost
  invariant sets are represented by the red and the blue regions at the ends
  of the eigenvector range.  One can see that the Fig.~\ref{fig:almostinvariant_control}
    left basin (red) is larger than the Fig.~\ref{fig:almostinvariant}(a) left
  basin, while the Fig.~\ref{fig:almostinvariant_control} right basin (blue)
  is smaller than the Fig.~\ref{fig:almostinvariant}(a) right basin.}
\end{figure}

The control scheme is tested by observing the
changes in the almost invariant sets.  Figure~\ref{fig:almostinvariant_control} shows the results of approximating
  the almost invariant sets for the stochastic system
with control applied to the left basin given by
Eqs.~(\ref{e:xdot_c})-(\ref{e:r_c}) for $c=0.25$. As expected, 
there is an increase in the size of the left (red)
almost invariant set and a decrease in the size of the
right (blue) almost invariant set when compared to
Fig.~\ref{fig:almostinvariant}. 

We have analyzed how individual trajectories escape under this control algorithm. By
measuring the most frequent path of an escape event from the left basin to the
right basin, one can observe that the trajectories follow the FTLE ridge
towards a periodic saddle orbit on the $x$-axis. Topologically, the FTLE ridge
loops around the almost invariant sets, corralling points in the outlying
region to the other side. The goal is to reduce the number of escape events by a control algorithm that takes advantage of the topology of the system.
 
\begin{figure}
\begin{center}
\begin{minipage}{0.49\linewidth}
\includegraphics[width=8.25cm]{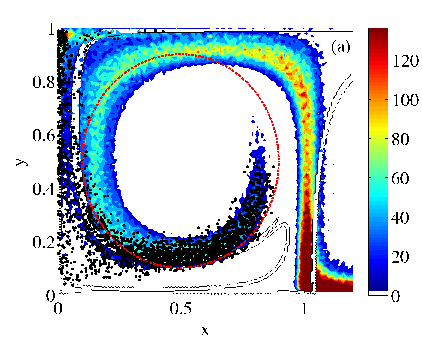}
\end{minipage}
\begin{minipage}{0.49\linewidth}
\includegraphics[width=8.25cm]{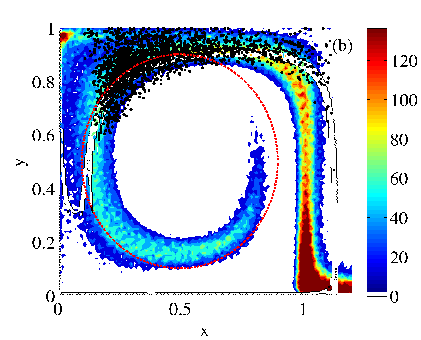}
\end{minipage}\\
\begin{minipage}{0.98\linewidth}
\includegraphics[width=8.25cm]{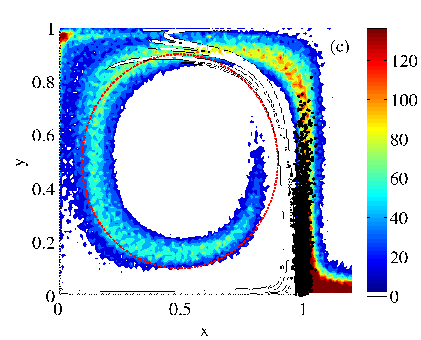} 
\end{minipage}
\caption{\label{fig:topology}(Color online) Probability density of the
  trajectory path prehistory for particles that escaped from the left basin
  based on a threshold of $x=1.2$ between phases $\psi\in [0.65,0.7)$. These
    paths form a subset of $20,000$ random particle simulations. Also shown is
    the position of the particles (black dots) before escape, the FTLE ridge (black curve), and the control region of radius $r=0.4$ (red) at (a) phase $\psi =0.0$, (b) phase $\psi =0.25$, and (c) phase $\psi =0.5$. The control force is $c=0.15$.}
\end{center}
\end{figure}

Figures~\ref{fig:topology}(a)-(c) illustrate the topology (as does a movie
which can be found as auxiliary online material). In these figures, $20,000$
particles were placed in the left basin near the fixed point and their
trajectories were tracked. If a particle crossed the $x=1.2$ threshold, the
particle was considered to have escaped from the left basin into the right
basin. In Figs.~\ref{fig:topology}(a)-(c), the color contours show the
probability density of the trajectory path prehistory for all the particles
(there are 4361 of them) that escaped from the left basin between phases
$\psi\in [0.65,0.7)$. The prehistory shows that there are many paths to
escape. However, the high probability density in the region of the saddle
(near the point $(1,0)$) shows that most of the particles have been funneled
into the saddle region just prior to escape from the left basin. Also shown in the figures are the position of the particles (black dots) before escape, the FTLE contour, and a circular region of radius $r=0.4$ that denotes the control region. These last three features are shown at phase $\psi=0.0$ in Fig.~\ref{fig:topology}(a), at phase $\psi =0.25$ in Fig.~\ref{fig:topology}(b), and at $\psi=0.5$ in Fig.~\ref{fig:topology}(c). The control force used in this simulation is $c=0.15$.

In Fig.~\ref{fig:topology}(a), one can see that although many particles are
outside the control region, the majority of the particles are inside the
control region. For those particles that are inside the control region, no
control is being applied. In addition, the majority of the particles lie on
one side of the FTLE ridge, which is bounding the almost invariant set (from
comparison with Fig.~\ref{fig:almostinvariant}). As time evolves, one can
see in Fig.~\ref{fig:topology}(b) that the FTLE ridge sweeps up the left side
of the figure, and one can see from the prehistory that about half of the
particles sweep up the left side, staying on the exterior side of the FTLE
ridge outside the control region so that the control is actuated.  The other half of the particles remain on the interior side of the FTLE ridge inside the control region so that no control is actuated.

Eventually, as can be seen in Fig.~\ref{fig:topology}(c), the FTLE ridge enters the region of uncertainty (see Fig.~\ref{fig:almostinvariant}), and all the particles cross over the FTLE ridge. All the particles now have left the control region. However, even though the control is being actuated, it cannot overcome the underlying dynamics found in this system and all the particles will proceed towards the saddle and finally escape. To prevent escape and to increase loitering time, one can use the knowledge of the location of the FTLE ridge, uncertainty regions and almost invariant sets to design a control region. For example, by choosing a control region with a smaller radius, one can prevent the particles from approaching the FTLE ridges and uncertain areas. With such a small radius, however, the number of control actuations will be very large. Control actuation may be optimized by using a control region with the largest possible radius so that the region does not intersect the FTLE structures. 

\subsection{Scaling of the  mean escape time with respect to noise and control}

We vary the control parameter, $c$, to study its effect on the mean escape
time.  One expects that an increase in $c$ will lead to a corresponding
increase in the mean time to escape.  Note that since the control force is finite, there is always a finite
  probability of escape even if control is actuated.  The control algorithm decreases the probability of trajectories
  visiting the regions that have a high probability of transition, and the
  mean time to escape reflects the decrease in PDF in the transition
  regions.

\begin{figure}
\begin{centering}
\includegraphics[width=8cm]{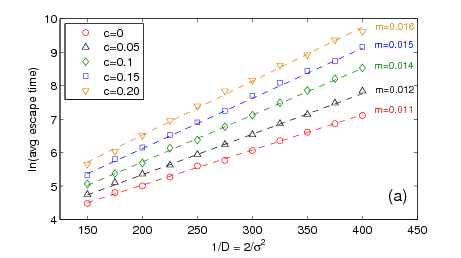} \includegraphics[width=6.5cm]{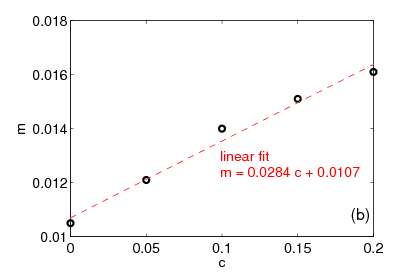} 
\par\end{centering}

\caption{\label{fig:escape} (Color online) (a) Natural log of the
  average escape time as a function of the inverse of the noise
  intensity.  The control radius is $r=0.4$ and a
  threshold of $x=1.2$ is used to identify escape from the left basin. The results are averaged over 1000 simulations and the slopes of the linear fit to each data set is noted on the right. (b) The slopes of the average escape time linear fits from Fig.~\ref{fig:escape}(a) as a function of the control parameter $c$.}

\end{figure}

Figure~\ref{fig:escape}(a) shows the natural log of the average escape time as
a function of the inverse of the noise intensity.  One can see that as the control force is increased, the mean time to escape increases with the slope depending linearly on the control parameter. Since small changes in $c$ translate into exponential changes in residence times, one recovers a linear relation of the slope of the exponent as a function of $c$ which can be seen in Fig.~\ref{fig:escape}(b).

\subsection{Optimizing control  actuations}

\begin{figure}
\begin{centering}
\includegraphics[width=8.5cm]{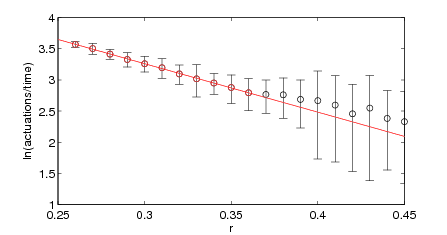} 
\par\end{centering}

\caption{\label{fig:actuate} Natural log of the average number of actuations per unit time as a function of the control radius for parameters $\sigma=0.075$ and $c=0.25$. The mean number of actuations $(\circ)$ was calculated over 100 simulations of time series with $t\le 10^5$, or until an escape event occurred. The error bars show one standard deviation of the data from the mean. Overlaid is a linear fit for the data points $r\le0.36$, since for this range of $r$ the control region does not intersect the FTLE ridge. }

\end{figure}

Another consideration is that one would like to minimize the number of
actuations for the particle, thereby preserving the energy needed to employ
the control scheme. As a consequence, this could extend the use of a battery
in an autonomous underwater glider. 
Figure~\ref{fig:actuate} shows the natural log of the average number of
actuations per unit time as a function of the control radius. For $r\le0.36$,
the FTLE ridge does not intersect the control region and the average number of
actuations follows the scaling of the average escape time. Beyond that, the number of actuations increases as the control scheme works to move the trajectory back inside the control region. Therefore, the number of actuations per unit time follows the average escape rate for $r\le0.36$ and the minimum number of actuations can be found for the largest control region inside the FTLE ridge.

\subsection{One-dimensional analysis of control}

To understand the effectiveness of the control scheme and its scaling properties, we approximate the
natural probability density function (PDF) and the almost invariant set in the
following 1D example.  Consider a stochastic system, where the deterministic
part has a stable focus at the origin surrounded by an unstable limit cycle.
In the associated deterministic system, the points inside the limit cycle
would spiral toward the stable focus, while the points outside would
diverge. The stochastic perturbations allow trajectories to escape
across the limit cycle (and then diverge), creating an almost invariant set. This captures the local behavior
in one basin of the double-gyre.  Now consider a control region with radius $\epsilon>0$, where $\epsilon$ determines the distance from the attractor for which the control is actuated. Then the system has the form 
\begin{equation}
\frac{d}{dt}\left[\begin{array}{c} x \\ y \end{array}\right] =
\left[\begin{array}{cc} -\lambda & \omega \\ - \omega & -\lambda\end{array}\right]
\left[\begin{array}{c} x \\ y \end{array}\right] + 
\lambda\left[\begin{array}{c} x \\ y \end{array}\right](x^{2}+y^{2}) - 
c\Theta(r-\epsilon) \left[\begin{array}{c} x \\ y \end{array}\right] + \bm{\eta}(t),
\label{eq:Unstable lc controller}
\end{equation}
where $\lambda,\omega>0$.  As before, $c$ is the magnitude of the
  control force and $\Theta$ is a Heaviside function.  Each
component of $\bm{\eta}(t)=\left[\eta_{1}(t),\eta_{2}(t)\right]^{\mbox{T}}$,
is a random variable with intensity $D$.

To find the PDF, we switch to polar coordinates using the time-dependent change of variables given
  by $x=r\cos\theta$ and $y=r\sin\theta$.  This results in the following transformed stochastic system: 
\begin{equation}
\dot{r}=\lambda r(r^{2}-1)-cr\Theta(r-\epsilon)+\eta_{1}(t)\cos\omega t -
\eta_{2}(t)\sin\omega t.
\label{eq:radial lc controller}
\end{equation}
Here, we find that $\theta=-\omega t$ and the noise vanishes entirely from the $\theta$ equation.

Assuming the transformed noise term still has intensity $D$, and $\kappa$ is a normalization constant, the probability density $\rho(r),$ is given by 
\begin{equation}
\rho(r)=\kappa\exp\left(\frac{\lambda}{4D}(r^{2}(r^{2}-2))-\frac{c}{2}(r^{2}-\epsilon^{2})\Theta(r-\epsilon)\right).
\label{eq:radial PDF}
\end{equation}
The PDF is now fully specified by the deterministic flow and noise drift, control strength and actuation region, and can be plotted for various parameters.

\begin{figure}
\begin{centering}
\includegraphics[width=8.5cm]{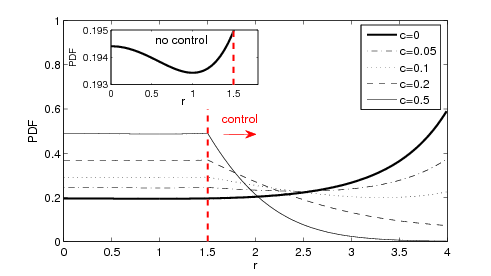} 
\par\end{centering}

\caption{\label{fig:PDF} (Color online) PDF as a function of the control
  radius $r$ for varying control force $c$ found using Eq.~(\ref{eq:radial
    PDF}).  Continuous noise is used with $\sigma=1$, $\lambda=0.1$, and
  $\epsilon=1.5$. The normalization constant, $\kappa$, was set in each case
  so that the area under the curve is one. The inset shows a close-up of the
  PDF near the attractor for the case of no control.}

\end{figure}

In Fig.~\ref{fig:PDF} one can see how an increase in the control strength decreases the probability of the trajectory moving outside the boundary of the control region. The inset figure shows that the local minimum of the PDF occurs at $r=1$, which is the location of the unstable limit cycle. The PDFs branch at the point of control at $r=1.5$. The thick solid curve represents the dynamics with no control. It follows that a trajectory escaping outside the limit cycle will diverge to infinity, and the PDF increases as one increases $r$.

\begin{figure}
\begin{centering}
\includegraphics[width=8cm]{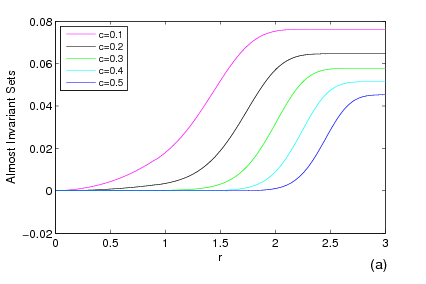} 
\includegraphics[width=8cm]{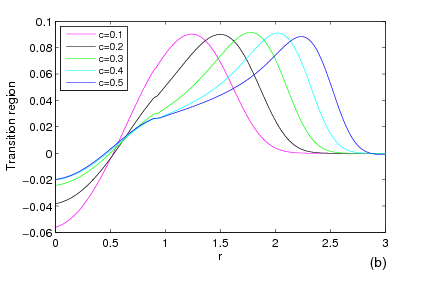} 
\par\end{centering}
\caption{\label{fig:AIS_control} (Color online) The results of the almost invariant sets analysis for varying control force $c$ in Eq.~(\ref{eq:radial lc controller}) with $\lambda=0.1$, $\epsilon=0.9$, and Gaussian noise with standard deviation of $\sigma = 0.1$. Figure~\ref{fig:AIS_control}(a) shows the almost invariant set for varying control force $c$. The second eigenvector of the transport matrix is mapped back to the domain $r$. Notice that as $c$ increases, the right basin extends to the right. Figure~\ref{fig:AIS_control}(b) shows the transition set for varying control force $c$. The third eigenvector of the transport matrix is mapped back to the domain $r$. Notice that as $c$ increases, the transition region moves to the right. }

\end{figure}

In addition, one can perform the one-dimensional almost invariant set analysis
for this example.  The transport matrix is computed using a grid of 500
intervals on the domain and Gaussian noise with standard deviation of $\sigma
= 0.1$. The point of control is located at $r=0.9$, to the left of the basin
boundary at $r=1$.  Figure~\ref{fig:AIS_control}(a), shows the expansion of
the left almost invariant set to the right as the control is increased. This
information is contained in the second eigenvector of the transport
matrix. The lower and upper almost constant regions of the function represent
location of the sets.  Figure~\ref{fig:AIS_control}(b), shows the movement of
the transport region to the right as the control is increased. This information is contained in the third eigenvector of the transport matrix. The maximum value of the function represents the location of the transition region. Notice that it is the complement of the almost invariant sets.

Both the PDF and almost invariant set analysis demonstrate the change that the control algorithm has on the natural dynamics of the stochastic system. The control algorithm decreases the probability of trajectories visiting regions where deterministic dynamics would cause them to diverge. This moves the effective basin boundary farther from the attractor, increasing its size. By using sufficient control radius and force, any trajectory that would naturally diverge can be redirected towards the attractor. Therefore, the almost invariant set can be expanded to the desired size at a cost of the number of control actuations.
 
Additionally, it is possible to quantify the relation between the mean
  escape time and the potential defined by the PDF.  Using the well-known Kramer's escape rate \cite{kra40}, one can predict the
rate at which a particle can escape over a potential barrier under Brownian
motion. In one dimension, the escape time of a particle from a potential
defined by $U(x)$ is
$\tau=\frac{2\pi}{\sqrt{U''(x_{a})|U''(x_{f})|}}\exp{[(U(x_{f})-U(x_{a}))/D]}$,
where $x_{a}$ is the location of the attractor and $x_{f}$ is the location of
the escape point. From the radial control PDF given by Eq.~(\ref{eq:radial
  PDF}), one can see that the potential function exponent depends linearly on
the control amplitude $c$. All of these one-dimensional analytic results are consistent with
the behavior seen for the double-gyre system with control that was presented in the preceding subsections.

\section{Conclusions}\label{sec:conc}

Using modern dynamical systems theory, we have analyzed the flow of a
stochastic double-gyre as a model of a wind-driven ocean. The existence of
multiple almost invariant sets combined with highly uncertain regions in phase
space gives rise to stochastic switching between the basins. The uncertain
regions generate high probability transition sets which are the primary cause
of switching from one basin to another. We showed that the high probability of
transitions may be approximated by computing uncertain regions with respect to
initial conditions and the use of FTLE. 

The system displays maximum sensitivity
near the deterministic basin boundary trajectory.  Knowledge of the FTLE ridges in a given flow, in conjunction with knowledge of uncertain regions and
almost invariant sets, enables one to predict the location of escape
trajectories from one basin to another.
As in previous work \cite{Schwartz2004}, these transition regions may be monitored directly to predict future switching events. 
This knowledge, in turn, allows for set-based corral control methods to be used to inhibit the escape event and increase the loitering time within the basin.

Almost invariant sets were computed explicitly using the Frobenius-Perron theory for discrete stochastic systems. The sets provided approximate target regions in which to control particles which act as sensors in large surveillance regions. That is, given initial conditions in one of the almost invariant basins, the particles will stay in the sets for long, but finite time. Although computed for a discrete map model, the sets approximate well the continuous noise almost invariant sets. They also form sets of the most certain points in that they are the complement of the uncertain region, which was computed based on initial condition uncertainty. 

Since the almost invariant sets are indicative of long-time, but finite, dynamics, we used these sets or subsets to define corral control regions in which we wish to maintain the residence times in one basin with minimal control actuation. To this end, we designed a simple discrete controller to extend the residence times in the continuous noise double-gyre. We find that even small control amplitudes have an exponential increase on the residence times. This is quite similar to the well-known problem of noise-induced escape from a potential well, even though the basin boundaries may be fractal in the deterministic case. 
 
We saw that if the control radius defined a set which did not intersect the ridges defined by the FTLE, the number of actuations per unit time followed an exponential law as a function of noise intensity. This can be understood completely from the switching rate theory. On the other hand, if the control radius intersected the FTLE ridge, the uncertainty regions were breached, and the control actuation rate no longer follows a nice scaling law, but rather some complicated function of control radius $r$. 

The model presented here is a simplified version of a double-gyre flow that is a
  solution to a realistic quasi-geostrophic ocean model, and the vehicles
controlled are point particles.  However, the techniques here can be extended to full ocean and glider models in the future. The extensions to vehicles with real mass means that inertial effects will need to be included, as well extensions to 3D GFD models. However, the machinery presented here is well-suited to the description of sets in higher dimensions, and we expect that the monitoring of large surveillance regions in the ocean by gliders will be enhanced by implementing our corral control method. Additionally, the study and control of coupled systems, with and without delay, is of interest. 

\section*{Acknowledgments}
We gratefully acknowledge support from the Office of Naval Research.  E.F. is
supported by the Naval Research Laboratory (Award No. N0017310-2-C007).

\begin{appendix}
\section{Details of the Ocean Model}\label{sec:ocean_model}
The ocean model results in Fig.~\ref{fig:ocean} were obtained by numerical solution of barotropic quasi-geostrophic flow in a single layer basin ${\bf \Omega}:=[0,1]\times[0,1]$ on a $\beta$-plane. The governing non-dimensionalized equation for the fluid streamfunction $\Psi$ is:
\begin{equation}\label{btqg3}
\frac{\partial\nabla^2\Psi}{\partial t} + \epsilon
J(\Psi,\nabla^2\Psi) + \frac{\partial\Psi}{\partial x}
= \mu \nabla^2\Psi + W ,
\end{equation}
where $J$ is the Jacobian operator,
\begin{equation}
J(f,g):=\frac{\partial f}{\partial x}\frac{\partial g}{\partial y} -
\frac{\partial g}{\partial x}\frac{\partial f}{\partial y},
\end{equation}
and forcing is provided by a wind stress curl, $W$, that is prescribed as
  follows to form a double-gyre circulation with a weak periodic ``seasonal'' variation:
\begin{equation}
W = -\sin{2\pi y} + 2\alpha\pi \sin{\omega t} ,
\end{equation}
where the amplitude $\alpha=0.1$ and frequency $\omega=3$ were used to produce Fig.~\ref{fig:ocean}.

This system is characterized by the non-dimensional parameters~\cite{Pedlosky2}
\begin{equation}
\mu = \frac{R}{\beta L} \;\;\; {\rm and} \;\;\; 
\epsilon = \frac{U}{\beta L^2}, 
\end{equation}
where $\beta$ is the rotation parameter, $R$ is the bottom friction and $L$
and $U$ are respectively
the characteristic length and velocity scales of the basin.
The parameters $\mu$ and $\epsilon$ correspond to the relative length scales of the Stommel and inertial
layers, respectively.  
To produce Fig.~\ref{fig:ocean}, we
  used $\mu=0.04$ and $\epsilon=0.0004$. 

 The above model has been numerically integrated using second-order spatial
 differences and second-order Runge-Kutta time stepping for
 the streamfunction, and a fourth-order Runge-Kutta algorithm to compute the Lagrangian trajectories of tracer particles. In Fig.~\ref{fig:ocean}, a grid of resolution $64^2$ was used and a dimensionless time step of $dt=0.001$. Coordinates of tracer particles are independent of the grid while flow velocities at the particle locations are found using bilinear interpolation from the grid values. An initially static ocean spins-up in a few hundred time steps. If $\alpha=0$ the spun-up solution is stationary, while non-zero $\alpha$ leads to a superimposed oscillatory behavior. The tracers are held in place until spin-up is complete.
\end{appendix}
 

\end{document}